\documentclass[12pt,a4paper]{article} 
\usepackage{amsfonts}
\usepackage{amsmath}
\newcommand{\mC}{{\mathbb C}}
\newcommand{\mR}{{\mathbb R}}

\title{Minimalization of uncertainty relations\\ in noncommutative quantum mechanics}
\author{Katarzyna Bolonek\\ Piotr Kosi\'nski\thanks{supported by KBN grant no 5P03B05620}\\
	 Department of Theoretical Physics II\\
	 University of \L{}\'od\'z\\
	 Pomorska 149/153, 90-236 \L{}\'od\'z\\
	 Poland}
\begin{document}
\maketitle
\begin{abstract}
\small
The explicit construction of states saturating uncertainty relations following 
from basic commutation rules of NCQM is given both in Fock space and coordinate representation.
\end{abstract}
\newpage

\section{Introduction}

There are strong indications coming from the study of brane configurations in string theory or 
matrix model of M-theory that noncommutative spaces are of some importance for very high 
energy physics \cite{1}. As a result, there appeared a large number of papers devoted  to
the study of field theories on such spaces \cite{2}. In order to reveal the important aspects of 
quantum theory on noncommutative spaces one should tend to simplify the systems under 
consideration as much as possible. By considering the low-energy limit of one-particle 
sector of field theory on noncommutative space one arrives at what is called noncommutative 
quantum mechanics. Again various aspects of it have been studied recently \cite{3}-\cite{23}.
In particular, in \cite{23} we considered single-particle quantum mechanics on noncommutative 
plane defined by the following commutation rules
\begin{subequations}
\label{w1}
\begin{align}
[\hat{x}_i,\hat{x}_j]&=i\theta\epsilon_{ij}I&\label{w1a}\\
[\hat{x}_i,\hat{p}_j]&=i\hbar\delta_{ij}I&i,j=1,2\label{w1b}\\
[\hat{p}_i,\hat{p}_j]&=0;&\label{w1c}
\end{align}
\end{subequations}
here we can assume $\theta>0$ without loosing generality.

By standard arguments, eqs.~(\ref{w1}) result in the following uncertainty relations
\begin{subequations}
\label{w2}
\begin{align}
\Delta x_1\Delta x_2&\geq\frac{\theta}{2}\label{w2a}\\
\Delta x_1\Delta p_1&\geq\frac{\hbar}{2}\label{w2b}\\
\Delta x_2\Delta p_2&\geq\frac{\hbar}{2}\label{w2c}
\end{align}
\end{subequations}
In the previous paper \cite{23} we studied the above inequalities in some detail.
In particular we have shown that, contrary to the commutative ($\theta=0$) case,
for a given state $\psi$ at most one of the inequalities~(\ref{w2}) can be saturated.
We have also outlined the construction of the states saturating any of them.

In the present paper we support and extend these results by explicit calculations.
In section~\ref{s2} we find (or, rather, remind) the construction of Fock space representation
of the algebra~(\ref{w1}). Then, in sec.~\ref{s3}, the explicit construction of all
states saturating the uncertainty relations~(\ref{w2}) is given; the relevant ingredients
here are the standard construction of coherent states and appropriate Bogolubov transformations.
Sec.~\ref{s4} is devoted to the study of minimalizing states in coordinate representation.
Their coordinate wave functions are given explicitely and it is checked by
straightforward calculations that no wave function exists which saturates more than one 
of the inequalities~(\ref{w2}). Finally, some basic facts concerning the standard coherent
states are collected in Appendix.

\section{Representations of the basic algebra}
\label{s2}

It is not difficult to find irreducible representation of the algebra~(\ref{w1}). 
In fact, this algebra is equivalent to standard Heisenberg-Weyl algebra:
\begin{equation}
\label{w3}
\begin{split}
\tilde{x}_i&\equiv\hat{x}_i+\frac{\theta}{2\hbar}\epsilon_{ij}\hat{p}_j\\
\tilde{p}_i&\equiv\hat{p}_i
\end{split}
\end{equation}
obey standard H-W commutation rules.
Eq.~(\ref{w3}) suggests the following definition of creation/anihilation operators 
(we work with $\omega=1$, $m=1$ units)
\begin{equation}
\label{w4}
\begin{split}
a_i&\equiv \frac{1}{\sqrt{2\hbar}}(\hat{x}_i+(i\delta_{ij}+\frac{\theta}{2\hbar}
\epsilon_{ij})\hat{p}_j)\\
a_i^\dagger&\equiv \frac{1}{\sqrt{2\hbar}}(\hat{x}_i+(-i\delta_{ij}+\frac{\theta}{2\hbar}
\epsilon_{ij})\hat{p}_j)
\end{split}
\end{equation}
Then the only nonvanishing commutator reads
\begin{equation}
\label{w5}
[a_i,a_j^\dagger]=\delta_{ij}
\end{equation}
and we arrive at the standard Fock space spanned by the orthonormal vectors
\begin{equation}
\label{w6}
|n_1,n_2\rangle=\frac{1}{\sqrt{n_1!}}\frac{1}{\sqrt{n_2!}}(a_1^\dagger)^{n_1}
(a_2^\dagger)^{n_2}|0\rangle
\end{equation}
The inverse relations to~(\ref{w4}) read 
\begin{equation}
\label{w7}
\begin{split}
\hat{x}_i&=\sqrt{\frac{\hbar}{2}}(a_i+a_i^\dagger)+
\frac{i\theta}{2\sqrt{2\hbar}}\epsilon_{ij}(a_j-a_j^\dagger)\\
\hat{p}_i&=\sqrt{\frac{\hbar}{2}}\frac{a_i-a_i^\dagger}{i}
\end{split}
\end{equation}
It is often convenient to work with the modified creation/anihilation operators 
carrying definite angular momentum. To this end we define \
\begin{equation}
\label{w8}
\begin{split}
a_{\pm}&\equiv\frac{1}{\sqrt{2}}(a_1\mp ia_2)\\
a_{\pm}^\dagger&\equiv\frac{1}{\sqrt{2}}(a_1^\dagger\pm ia_2^\dagger);
\end{split}
\end{equation} 
The new basis is 
\begin{equation}
\label{w9} 
|n_{+},n_{-}\rangle=\frac{1}{\sqrt{n_{+}!}}\frac{1}{\sqrt{n_{-}!}}(a_{+}^\dagger)^{n_{+}}
(a_{-}^\dagger)^{n_{-}}|0\rangle
\end{equation}
In terms of new variables the angular momentum operator reads
\begin{equation}
\label{w10}
\hat{L}=-i\hbar\epsilon_{ij}a_i^\dagger a_j=\hbar(a_{+}^\dagger a_{+}-a_{-}^\dagger a_{-})
\end{equation}
The angular momentum of the state~(\ref{w9}) equals $\hbar(n_{+}-n_{-})$.

\section{Saturating uncertainty relations}
\label{s3}
Let us first find all vectors saturating the uncertainty relation~(\ref{w2a}). The relevant
commutation rule~(\ref{w1a}) resembles the one concerning $\hat{x}_1$ and $\hat{p}_1$,
with $\hat{p}_1$ replaced by $\hat{x}_2$ and $\hbar$ replaced by $\theta$. Therefore,
it is not surprising that we can use the same strategy as described in Appendix once
the appropriate creation/anihilation operators are found. To this end we define
\begin{equation}
\label{w11}
\begin{split}
b&\equiv\sqrt{\frac{\hbar}{2\theta}}\left(
\left(1+\frac{\theta}{2\hbar}\right)a_{-}+
\left(1-\frac{\theta}{2\hbar}\right)a_{+}^\dagger\right)\\
b^\dagger&\equiv\sqrt{\frac{\hbar}{2\theta}}\left(
\left(1+\frac{\theta}{2\hbar}\right)a_{-}^\dagger+
\left(1-\frac{\theta}{2\hbar}\right)a_{+}\right)\\
c&\equiv\sqrt{\frac{\hbar}{2\theta}}\left(
\left(1+\frac{\theta}{2\hbar}\right)a_{+}+
\left(1-\frac{\theta}{2\hbar}\right)a_{-}^\dagger\right)\\
c^\dagger&\equiv\sqrt{\frac{\hbar}{2\theta}}\left(
\left(1+\frac{\theta}{2\hbar}\right)a_{+}^\dagger+
\left(1-\frac{\theta}{2\hbar}\right)a_{-}\right)
\end{split}
\end{equation}
One easily verifies that $b$, $c$, $b^\dagger$, $c^\dagger$ form the set of independent
creation/ani\-hi\-la\-tion operators.

The key point is that $b$--operators are related to $\hat{x}$--operators in the standard way
\begin{equation}
\label{w12}
\begin{split}
b&\equiv\frac{1}{\sqrt{2\theta}}(\hat{x}_1+i\hat{x}_2)\\
b^\dagger&\equiv\frac{1}{\sqrt{2\theta}}(\hat{x}_1-i\hat{x}_2)
\end{split}
\end{equation}
Therefore, we can repeat the procedure outlined in Appendix to find the states saturating 
(\ref{w2a}). They read
\begin{equation}
\label{w13}
|z,\gamma\rangle_\phi=e^{-\frac{1}{2}|z|^2}e^{+\frac{1}{4}\ln\gamma((b^\dagger)^2-b^2)}
e^{zb^\dagger}|\phi\rangle
\end{equation}
where $|\phi\rangle$ is arbitrary state such that 
\begin{equation}
\label{w14}
b|\phi\rangle=0
\end{equation}
The ''vacuum'' state is by far not unique;
it may contain an arbitrary number of $c$--excitations.

The representation given by $b$, $b^\dagger$, $c$, $c^\dagger$ is unitary equivalent to
that defined by $a_\pm$, $a_\pm^\dagger$. In fact, one can check that
\begin{equation}
\label{w15}
\begin{aligned}
b&=Wa_{-}W^\dagger,&b^\dagger&=Wa^\dagger_{-}W^\dagger\\
c&=Wa_{+}W^\dagger,&c^\dagger&=Wa^\dagger_{+}W^\dagger
\end{aligned}
\end{equation}
where
\begin{equation}
\label{w16}
W=e^{\frac{1}{2}\ln\left(\frac{2\hbar}{\theta}\right)(a_{+}a_{-}-a_{+}^\dagger a_{-}^\dagger)}
\end{equation}
This can be seen by using the results of \cite{23}. However, we prefer to give a
straightforward proof. Define for any $t\in\mR$
\begin{equation}
\label{w17}
W(t)=e^{t(a_{+}a_{-}-a_{+}^\dagger a_{-}^\dagger)}
\end{equation}
and 
\begin{equation}
\label{w18}
\begin{split}
b(t)&\equiv W(t)a_{-}W^\dagger(t)\\
c^\dagger(t)&\equiv W(t)a_{+}^\dagger W^\dagger(t)
\end{split}
\end{equation}
Then $b(0)=a_{-}$, $c^\dagger(0)=a_{+}^\dagger$ while simple computation gives
\begin{equation}
\label{w19}
\left(\begin{array}{l}
\dot{b}(t)\\
\dot{c}^\dagger(t)
\end{array}\right)=
\left(\begin{array}{ll}
0&1\\
1&0
\end{array}\right)
\left(\begin{array}{l}
b(t)\\
c^\dagger(t)
\end{array}\right)
\end{equation}
Therefore
\begin{equation}
\label{w20}
\begin{split}
\left(\begin{array}{l}
b(t)\\
c^\dagger(t)
\end{array}\right)&=
\left(\exp\left(\begin{array}{ll}
0&t\\
t&0
\end{array}\right)\right)
\left(\begin{array}{l}
a_{-}\\
a_{+}^\dagger
\end{array}\right)={}\\
&=\left(\begin{array}{ll}
\cosh t & \sinh t\\
\sinh t & \cosh t
\end{array}\right)
\left(\begin{array}{l}
a_{-}\\
a_{+}^\dagger
\end{array}\right)
\end{split}
\end{equation}
For $t=\frac{1}{2}\ln\left(\frac{2\hbar}{\theta}\right)$ we arrive at (\ref{w15}).

Eqs.~(\ref{w15}), together with the results presented in Appendix allow us 
to conclude that the states saturating (\ref{w2a}) are linear combinations
(with respect to $n_{+}$ but with $z$, $\gamma$ fixed) of the states

\begin{equation}
\label{w21}
|z,\gamma,n_{+}\rangle=e^{-\frac{1}{2}|z|^2}W
e^{-\frac{1}{4}\ln\gamma(a_{-}^2-(a_{-}^\dagger)^2)}
e^{za_{-}^\dagger}|n_{+},0\rangle
\end{equation}
Let us note that $W$ commutes with $\hat{L}$. This implies that the states $z=0$,
$\gamma=1$ are eigenstates of $\hat{L}$. This conclusion is rather obvious:
real and imaginary parts of $z$ are related to expectation values of $\hat{x}_1$,
$\hat{x}_2$ (which should be zero from rotational invariance) while expectation values
of $\hat{x}_1^2$, resp. $\hat{x}_2^2$ are proportional to $\gamma$, resp. $\frac{1}{\gamma}$.

Let us now consider the states saturating 
\begin{equation}
\label{w22}
\Delta x_1\Delta p_1\geq \frac{\hbar}{2}
\end{equation}
We follow the same strategy. First, define new creation/anihilation operators
\begin{equation}
\label{w23}
\begin{split}
d&=a_1+\frac{i\theta}{4\hbar}(a_2-a_2^\dagger)\\
d^\dagger&=a^\dagger_1+\frac{i\theta}{4\hbar}(a_2-a_2^\dagger)\\
e&=a_2+\frac{i\theta}{4\hbar}(a_1-a_1^\dagger)\\
e^\dagger&=a_2^\dagger+\frac{i\theta}{4\hbar}(a_1-a_1^\dagger)
\end{split}
\end{equation}
which obey
\begin{equation}
\label{w24}
d=\frac{1}{\sqrt{2\hbar}}(\hat{x}_1+i\hat{p}_1)
\end{equation}
Unitary equivalence of old and new operators,
\begin{equation}
\label{w25}
\begin{split}
d&=Ta_1T^\dagger\\
e&=Ta_2T^\dagger
\end{split}
\end{equation}
is obtained by choosing T in the form (cf. \cite{23})
\begin{equation}
\label{w26}
T=e^{\frac{i\theta}{4\hbar}(a_1-a_1^\dagger)(a_2-a_2^\dagger)}
\end{equation}
Consequently the states saturating (\ref{w26}) can be written as linear combinations,
with respect to $n_2$ but with $z$, $\gamma$ fixed, of the states
\begin{equation}
\label{w27}
|z,\gamma,n_2\rangle=e^{-\frac{1}{2}|z|^2}Te^{-\frac{1}{4}\ln\gamma(a_1^2-(a_1^\dagger)^2)}
e^{za_1^\dagger}|0,n_2\rangle
\end{equation}
The states saturating (\ref{w2c}) are obtained by replacing $1\leftrightarrow2$,
$\theta\rightarrow-\theta$:
\begin{equation}
\label{w28}
|z,\gamma,n_1\rangle=e^{-\frac{1}{2}|z|^2}T^\dagger 
e^{-\frac{1}{4}\ln\gamma(a_2^2-(a_2^\dagger)^2)}
e^{za_2^\dagger}|0,n_1\rangle
\end{equation}

\section{Coordinate representation}
\label{s4}

For the variables $\tilde{x}_i$, $\tilde{p}_i$ we use standard representation
\begin{equation}
\label{w29}
\begin{split}
\tilde{x}_i&=x_i\\
\tilde{p}_i&=-i\hbar\frac{\partial}{\partial x_i}
\end{split}
\end{equation}
which implies
\begin{equation}
\label{w30}
\begin{split}
\hat{x}_i&=x_i+\frac{i\theta}{2}\epsilon_{ij}\frac{\partial}{\partial x_j}\\
\hat{p}_i&=-i\hbar\frac{\partial}{\partial x_i}
\end{split}
\end{equation}
The state $\psi$ saturating (\ref{w2a}) obeys
\begin{equation}
\label{w31}
(\hat{x}_1-\alpha)\psi=-i\gamma(\hat{x}_2-\beta)\psi
\end{equation}
which , due to eqs.~(\ref{w30}), takes the form
\begin{equation}
\label{w32}
\frac{\theta}{2}\left(\gamma\frac{\partial}{\partial x_1}+i\frac{\partial}{\partial x_2}\right)
\psi+((x_1+i\gamma x_2)-(\alpha+i\gamma\beta))\psi=0
\end{equation}
The general solution reads
\begin{equation}
\label{w33}
\psi(x_1,x_2)=f\left(\frac{x_1}{\sqrt{\gamma}}+i\sqrt{\gamma}x_2\right)
e^{-\frac{1}{\theta}\left(\left(\frac{x_1^2}{\gamma}+\gamma x_2^2\right)-
z \left(\frac{x_1}{\sqrt{\gamma}}-i\sqrt{\gamma}x_2\right)\right)}
\end{equation}
with $z\equiv\frac{\alpha}{\sqrt{\gamma}}+i\sqrt{\gamma}\beta$; f is an arbitrary function
such that $\psi$ is normalizable. In particular, the eigenstate of $\hat{L}$ corresponding
to the eigenvalue $\hbar m$ reads
\begin{equation}
\label{w34}
\psi(x_1,x_2)=\frac{2^{\frac{m+1}{2}}}{\sqrt{\pi}\sqrt{m!}\theta^{\frac{m+1}{2}}}
e^{im\phi}r^me^{-\frac{r^2}{\theta}}
\end{equation}
One can check explicitly that 
$\langle \hat{x}^2_1\rangle=\langle\hat{x}_2^2\rangle=\frac{\theta}{2}$ as it should be.

Let us note that only eigenstates with nonnegative eigenvalues $m\geq 0$ can saturate 
(\ref{w2a}). This can be easily understood. We have
\begin{equation}
\label{w35}
\langle\hat{x}_1^2\rangle=\langle\hat{x}_2^2\rangle=\frac{1}{2}
\langle \hat{x}_1^2+\hat{x}_2^2\rangle=
\frac{1}{2}\langle\tilde{x}_1^2+\tilde{x}_2^2+\frac{\theta^2}{4\hbar^2}
(\tilde{p}_1^2+\tilde{p}_2^2)-\frac{\theta}{\hbar}\hat{L}\rangle;
\end{equation}
the right-hand side is the combination of harmonic oscillator and angular momentum.
Standard reasoning gives for the spectra
\begin{equation}
\label{w36}
\begin{split}
\tilde{x}_1^2+\tilde{x}_2^2+\frac{\theta^2}{4\hbar^2}(\tilde{p}_1^2+\tilde{p}_2^2)
-\frac{\theta}{\hbar}\hat{L}\colon\ \ &\theta(2n_{-}+1)\\
\hat{L}\colon\ \ &\hbar(n_{+}-n_{-})
\end{split}
\end{equation}
The states saturating~(\ref{w2a}) correspond to $n_{-}=0$;
but $n_{+}-n_{-}=m$, i.e. $m=n_{+}\geq 0$.

Let us look for the states saturating (\ref{w2a}). The relevant equation
\begin{equation}
\label{w37}
\hat{x}_1\psi_1=-i\gamma_1\hat{p}_1\psi_1
\end{equation}
reads
\begin{equation}
\label{w38}
\left(\frac{i\theta}{2}\frac{\partial}{\partial x_2}+\gamma_1\hbar\frac{\partial}{\partial x_1}
\right)\psi_1+x_1\psi_1=0
\end{equation}
and gives
\begin{equation}
\label{w39}
\psi_1=f_1(x_1+\frac{2i\gamma_1\hbar}{\theta}x_2)e^{
-\frac{3x_1^2}{8\gamma_1\hbar}-\frac{\gamma_1\hbar x_2^2}{2\theta^2}+
\frac{ix_1x_2}{2\theta}},
\end{equation}
where $f_1$ is arbitrary  such that $\psi$ is normalizable. 

The states saturating (\ref{w2c}) are obtained by replacement $x_1\leftrightarrow x_2$, 
$\theta\rightarrow-\theta$, $\gamma_1\rightarrow\gamma_2$
\begin{equation}
\label{w40}
\psi_2=f_2(x_2-\frac{2i\gamma_2\hbar}{\theta}x_1)e^{
-\frac{3x_2^2}{8\gamma_2\hbar}-\frac{\gamma_2\hbar x_1^2}{2\theta^2}-
\frac{ix_1x_2}{2\theta}},
\end{equation}
It is not difficult to show that there exists no state saturating both (\ref{w2b})
and (\ref{w2c}). To this end we insert (\ref{w40}) into eq.~(\ref{w37}) and find
\begin{equation}
\label{w41}
\frac{{f'}_2\left(x_2-\frac{2i\gamma_2\hbar}{\theta}x_1\right)}{
f_2\left(x_2-\frac{2i\gamma_2\hbar}{\theta}x_1\right)}=
\frac{\left(\frac{5}{4}-\frac{\gamma_1\gamma_2\hbar^2}{\theta^2}\right)x_1
-i\left(\frac{3\theta}{8\gamma_2\hbar}+\frac{\gamma_1\hbar}{2\theta}\right)x_2}
{\frac{2i\gamma_1\gamma_2\hbar^2}{\theta}-\frac{i\theta}{2}}
\end{equation}
The left-hand side depends only on one variable
$x_2-\frac{2i\gamma_2\hbar}{\theta}x_1$ so the right-hand side must also;
this is, however, imposible as one can immedietely check.

One can also ask whether (\ref{w39}) ((\ref{w40})) can be an eigenstate of $\hat{L}$ 
provided an appropriate choice of $f_1$ ($f_2$) has been made. Again we check that
this is impossible inserting (\ref{w39}) into the eigenequation
\begin{equation}
\label{w42}
\hat{L}\psi=\hbar m \psi
\end{equation}
Let us finally insert eq.~(\ref{w33}) into eq.~(\ref{w37}).
The resulting equation for the function $f$ reads
\begin{equation}
\label{w43}
\frac{f'\left(\frac{x_1}{\sqrt{\gamma}}+i\sqrt{\gamma}x_2\right)}
{f\left(\frac{x_1}{\sqrt{\gamma}}+i\sqrt{\gamma}x_2\right)}=
\frac{\left(1-\frac{2\gamma_1\hbar}{\gamma\theta}\right)x_1-i\gamma x_2+
z\left(\frac{\sqrt{\gamma}}{2}+\frac{\gamma_1\hbar}{\theta\sqrt{\gamma}}\right)}
{\frac{\theta\sqrt{\gamma}}{2}-\frac{\gamma_1\hbar}{\sqrt{\gamma}}};
\end{equation}
the consistency  condition (the right-hand side should depend only on 
$\frac{x_1}{\sqrt{\gamma}}+i\sqrt{\gamma}x_2$) implies
\begin{equation}
\label{w44}
\frac{\gamma_1\hbar}{\gamma\theta}=1
\end{equation}
Under this condition the solution to (\ref{w43}) reads
\begin{equation}
\label{w45}
f=Ce^{\frac{1}{\theta}
\left(\frac{x_1}{\sqrt{\gamma}}+i\sqrt{\gamma}x_2\right)^2
-\frac{3\sqrt{\gamma}}{\theta}z
\left(\frac{x_1}{\sqrt{\gamma}}+i\sqrt{\gamma}x_2\right)}
\end{equation}
Inserting this back to (\ref{w33}) we conclude that $\psi$ is nonnormalizable. This shows
that also (\ref{w2a}) and (\ref{w2b}) cannot be simultaneously saturated.

We verified explicitly that, for a given state $\psi$, at most one of the inequalities
(\ref{w2a})--(\ref{w2c}) can be saturated; this confirms the general theorems of \cite{23}.

Although there are no states saturating both (\ref{w2b}) and (\ref{w2c}),
both lower bounds can be simultaneously approached as close as one wishes. To see this 
we select the state
\begin{equation}
\label{w46}
\psi=\sqrt{\frac{2\delta}{\pi}}e^{-\delta(x_1^2+x_2^2)}
\end{equation}
Then $\hat{L}\psi=0$, $\langle\hat{p}_1\rangle=0$, $\langle\hat{x}_1\rangle=0$, and
\begin{align}
\langle\hat{p}_1^2\rangle_\psi&=\delta\hbar^2,&\langle\hat{x}_1^2\rangle&=
\frac{1}{4\delta}+\frac{\theta^2\delta}{4};
\label{w47}
\end{align}
consequently
\begin{equation}
\label{w48}
(\Delta x_1)^2_\psi(\Delta p_1)_\psi^2=\frac{\hbar^2}{4}+\frac{\theta^2\hbar^2\delta^2}{4}
\end{equation}
By symmetry
\begin{equation}
\label{w49}
(\Delta x_2)^2_\psi(\Delta p_2)_\psi^2=\frac{\hbar^2}{4}+\frac{\theta^2\hbar^2\delta^2}{4}.
\end{equation}
(\ref{w46}) is normalizable for any $\delta>0$. The bounds are saturated for 
$\delta\rightarrow 0$; however, the state (\ref{w46}) becomes nonnormalizable in the limit
$\delta\rightarrow 0$.

\newpage
\appendix
\section*{ Appendix: Uncertainty principles and coherent states}

First let us remind the general setting for uncertainty principles \cite{24}
(for recent alternative approach see \cite{25}). Given
two observables $\hat{A}$, $\hat{B}$ subject to commutation rule:
\begin{equation}
\label{wA1}
[\hat{A},\hat{B}]=i\hat{C},
\end{equation}
one can derive the following inequality (generalized Heisenberg principle)
\begin{equation}
\label{wA2}
(\Delta A)_\psi\cdot(\Delta B)_\psi\geq\frac{1}{2}|\langle C\rangle_\psi|,
\end{equation}
with $|\psi\rangle$ normalized to unity and
\begin{equation}
\label{wA3}
(\Delta A)_\psi=\sqrt{\langle\psi|(\hat{A}-\langle\hat{A}\rangle_\psi I)^2|\psi\rangle},\ \ \ 
\text{etc.};
\end{equation}
(\ref{wA2}) is saturated iff the following condition holds
\begin{align}
\label{wA4}
(\hat{A}-\langle\hat{A}\rangle_\psi I)|\psi\rangle=-i\gamma(\hat{B}-\langle B\rangle_\psi I)
|\psi\rangle, &\ \  \gamma\in\mR
\end{align}
Acting with $\hat{A}-\langle\hat{A}\rangle_\psi I$ on both sides of (\ref{wA4}), using (\ref{wA1})
and again (\ref{wA4}) one arrives at 
\begin{equation}
\label{wA5}
(\hat{A}-\langle\hat{A}\rangle_\psi I)^2|\psi\rangle=-\gamma^2(\hat{B}-\langle\hat{B}\rangle_\psi I)^2
|\psi\rangle+\gamma\hat{C}|\psi\rangle
\end{equation}
or, on multiplying by $|\psi\rangle$ from the left
\begin{equation}
\label{wA6}
(\Delta A)^2_\psi+\gamma^2(\Delta B)^2_\psi=\gamma\langle C\rangle_\psi.
\end{equation}
(\ref{wA6}), together with the saturated form of (\ref{wA2}) gives (provided $\gamma\neq 0$)
\begin{align}
\label{wA7}
(\Delta A)^2_\psi&=\frac{\gamma}{2}\langle C \rangle_\psi, &
(\Delta B)^2_\psi&=\frac{1}{2\gamma}\langle C\rangle_\psi
\end{align}
which explains the meaning of $\gamma$.

Let us apply this scheme to the standard Heisenberg relation
\begin{equation}
\label{wA8}
[\hat{x},\hat{p}]=i\hbar
\end{equation} 
The relevant inequality reads
\begin{equation}
\label{wA9}
\Delta x\cdot\Delta p\geq\frac{\hbar}{2};
\end{equation}
(\ref{wA9}) is saturated iff
\begin{align}
\label{wA10}
(\hat{x}-\alpha)|\psi\rangle&=-i\gamma(\hat{p}-\beta)|\psi\rangle,&
\alpha=\langle \hat{x} \rangle_\psi,\ \beta=\langle \hat{p}\rangle_\psi
\end{align}
Let us define creation/anihilation operators (we work with $\omega=1$, $m=1$ units)
\begin{equation}
\label{wA11}
\begin{split}
a&\equiv\frac{1}{\sqrt{2\hbar}}(\hat{x}+i\hat{p})\\
a^\dagger&\equiv\frac{1}{\sqrt{2\hbar}}(\hat{x}-i\hat{p})\\
{}[a,\lefteqn{a^\dagger]=1.}&
\end{split}
\end{equation}
Hilbert space of states is spanned by the vectors 
\begin{equation}
\label{wA12}
|n\rangle=\frac{1}{\sqrt{n!}}(a^\dagger)^n|0\rangle
\end{equation}
To find the general solution to (\ref{wA10}) first note that $\gamma>0$. In fact, $\gamma\neq 0$
because $\hat{x}-\alpha I$ cannot have normalized eigenvectors (operators commuting to 
$\mC$--number have no normalized eigenvectors in their common invariant domain); for $\gamma\neq 0$
(\ref{wA7}) gives $\gamma>0$. We start with $\gamma=1$. Eq.~(\ref{wA10}) can be rewritten as
\begin{align}
\label{wA13}
a|\psi\rangle&=z|\psi\rangle,&z&=\frac{\alpha+i\beta}{\sqrt{2\hbar}}
\end{align} 
The eigenstates of the anihilation operators are called coherent states (cs).
Vacuum state is the coherent state corresponding to $z=0$.
In order to find other cs one defines, for any $z\in\mC$, the unitary operators
\begin{equation}
\label{wA14}
U(z)\equiv e^{za^\dagger-\bar{z}a}=e^{-\frac{1}{2}|z|^2}e^{za^\dagger}e^{-\bar{z}a}
\end{equation}
One easily checks that 
\begin{equation}
\label{wA15}
U^\dagger(z)aU(z)=a+z\cdot I
\end{equation}
Therefore, the coherent states are given by 
\begin{equation}
\label{wA16}
|z\rangle\equiv U(z)|0\rangle=e^{-\frac{1}{2}|z|^2}e^{za^\dagger}|0\rangle=
e^{-\frac{1}{2}|z|^2}\sum_{n=0}^\infty \frac{z^n}{\sqrt{n!}}|n\rangle
\end{equation}
Consider now the case $\gamma\neq 1$. Eq.~(\ref{wA10}) can be written as
\begin{equation}
\label{wA17}
a_\gamma|\psi\rangle=z|\psi\rangle
\end{equation}
where
\begin{equation}
\label{wA18}
\begin{split}
z&=\frac{1}{\sqrt{2\hbar}}\left(\frac{\alpha}{\sqrt{\gamma}}+i\beta\sqrt{\gamma}\right)\\
a_\gamma&=\frac{1}{\sqrt{2\hbar}}\left(\frac{\hat{x}}{\sqrt{\gamma}}+i\sqrt{\gamma}\hat{p}\right)\\
a_\gamma^\dagger&=\frac{1}{\sqrt{2\hbar}}\left(\frac{\hat{x}}{\sqrt{\gamma}}-i\sqrt{\gamma}\hat{p}\right)
\end{split}
\end{equation}
Again, $[a_\gamma, a_\gamma^\dagger]=1$ and $a_{\gamma=1}=a$. Solutions to (\ref{wA17}) can be constructed
with the help of $a_\gamma$, $a_\gamma^\dagger$, and $\gamma$--vacuum $|0\rangle_\gamma$. However,
all representations of Fock algebra are unitarily equivalent. Indeed one can easily verify that,
with the unitary operator  $V(\gamma)$ defined by
\begin{equation}
\label{wA19}
V(\gamma)=e^{-\frac{1}{4}\ln\gamma(a^2-(a^\dagger)^2)},
\end{equation}
the following relations are obeyed
\begin{equation}
\label{wA20}
\begin{split}
V(\gamma)aV^\dagger(\gamma)&=a_\gamma\\
V(\gamma)a^\dagger V^\dagger(\gamma)&=a_\gamma^\dagger
\end{split}
\end{equation}
The solution to eq.~(\ref{wA10}) can be now written as 
\begin{equation}
\label{wA21}
|z,\gamma\rangle=V(\gamma)U(z)|0\rangle;
\end{equation}
the complex parameter $z$ is related to the mean values of $x$ and $p$ while $\gamma$ describes their
dispersions:
\begin{equation}
\label{wA22}
\begin{split}
(\Delta x)^2&=\frac{\gamma\hbar}{2}\\
(\Delta p)^2&=\frac{\hbar}{2\gamma}
\end{split}
\end{equation}

\end{document}